\documentclass[traditabstract,rnote]{aa} % for the abstract without structuration 
\usepackage{graphicx}
\usepackage{txfonts}
\def\la{\raise.5ex\hbox{$<$}\kern-.8em\lower 1mm\hbox{$\sim$}}
\def\ma{\raise.5ex\hbox{$>$}\kern-.8em\lower 1mm\hbox{$\sim$}}

\def\kms{$\rm km\, s^{-1}$}
\def\cm3{$\rm cm^{-3}$}
\def\Ts{$\rm T_{*}$~}
\def\Vs{$\rm V_{s}$~}
\def\n0{$\rm n_{0}$}
\def\B0{$\rm B_{0}$}

\def\erg{$\rm erg\, cm^{-2}\, s^{-1}$}
\def\mum{$\mu$m~}

\def\L12{L$_{12\mu m}$~}
\def\F12{F$_{12\mu m}$~}

\def\Hb{H${\beta}$~}
\def\Ha{H${\alpha}$~}

\def\Ly{Ly$\alpha$~}
\def\La{L$_{H\alpha}$~}
\def\Moy{M$_{\odot}$ yr$^{-1}$}

\begin{document}
   \title{Activity and quiescence  in  galaxies at redshifts 1.4$<$z$<$3.5.\\
The role of the starburst temperature.
}

   \author{M. Contini \inst{1,2}
}

   \institute{Dipartimento di Fisica e Astronomia, University of Padova, Vicolo dell'Osservatorio 2. I-35133 Padova, Italy
         \and
             School of Physics and Astronomy, Tel Aviv University, Tel Aviv 69978, Israel\\
%             \email{contini@post.tau.ac.il}
}

   \date{Received }

% \abstract{}{}{}{}{} 
% 5 {} token are mandatory
 
  \abstract{
We investigate  `activity' and `quiescence'  in galaxies at relatively high redshifts by modelling the line 
(and continuum) spectra  of each object.  The models account consistently for photoionization and shocks.
We claim that the  starburst effective temperature, the flux from an AGN, and the shock velocity are 
critical to activity.
The results confirm that  two sample galaxies show intense  starburst  activity with temperatures 
reaching \Ts=2 10$^5$K  and shock velocities \Vs$>$ 250 \kms, while for the remaining galaxies in our sample,
the models show quiescent star formation with \Ts $\leq$7 10$^4$K. 
A Seyfert 2 -like AGN is  proposed in one galaxy.  
The O/H relative abundances derived by the detailed modelling of the spectra are  nearly solar for all the sample galaxies,
in contrast to those obtained by direct methods.
}


   \maketitle

\section{Introduction}

In  previous papers (Contini 2014, hereafter Paper I, Contini 2013, and references therein) we have 
investigated the physical conditions and element abundances in galaxies   throughout  redshifts
0.001$<$z$<$3.4   by   detailed modelling of the emission line ratios.
We have found  that both active galactic nuclei (AGN) and starbursts (SB) coexist in  almost all galaxies and that
collisional process   should be accounted for.

In this paper we present  modelling  results  of 
 new spectra observed from selected  galaxies at 1.4$<$z$<$3.5 (Table 1), in particular those
with enough lines (e.g. of  oxygen in different ionization levels)  that  constrain the models  (Paper I). 
We cross-check the models, when possible, by the upper limits of significant lines,
by reproduction of the observed spectral energy distribution (SED) of the continuum,  
by the morphological evidence, etc. 

\begin{table}
\caption{Description of the observed galaxies}
\tiny{
\begin{tabular}{lccccccccccc} \hline  \hline
\ object    & z     & galaxy type   & telescope   &   ref \\  \hline
\ CDFS-695  & 2.225 & luminous  red  & Gemini NIRS&1 \\
\ 1255-0    & 2.1865&quiescent & Gemini NIRS&2   \\
\ A1835     & 2.073 &low-mass SF&Hale tripelspec&3 \\
\ A773      & 2.303 &low-mass SF&Hale tripelspec&3 \\
\ MACS0717  &2.55  &low-mass SF&Hale tripelspec&3 \\
\ WISP     &1.444-2.315&line selected      & Mag. Baade (FIRE)&4\\ 
\ J1000+0221S&3.417 & SF dwarf &CANDLES team& 5 \\ \hline
\end{tabular}}

1: van Dokkum et al (2005);2:  Kriek et al (2006);3: Belli et al (2013);4: Masters et al (2014);5: Amor\'{i}n et al. (2014) 
\end{table}

\begin{figure}
\includegraphics[width=10.2cm]{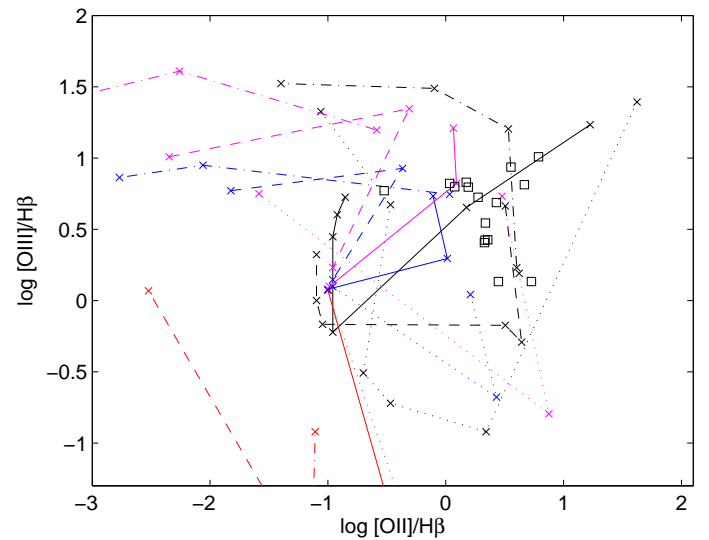}
\caption{
Black open squares: the observation data. Black: AGN(V=100 (dash-dot), 200 (dot), 300 (solid), 500 km s$^{-1}$ (dash);
x: $F$=0, log$F$=9,10,11,11.5,12. SB:red (T=10$^4$K), blue (T=5 10$^4$K), magenta (T=10$^5$K),
x:V=100,300,500\kms, $U$=0.01 (dot), 0.1 (solid), 1 (dash), 10 (dot-dash) 
}
\end{figure}

We associate the `activity' in galaxies (e.g. CDFS-695,  van Dokkum et al 2005) to
 an AGN, a SB, and to shocks on the basis of  a strong wind throughout the galaxy, while we refer to
`quiescent' galaxies e.g. 1255-0,   which  was classified  as a quiescent ultra-dense object by Kriek et al. (2009)
because of emitting  weak lines. Those imply a low luminosity AGN and 
 a low star formation rate (SFR = 1-3 \Moy)  compared with models of  stellar population
synthesis SEDs.   
In this paper we investigate activity and quiescence  in
 the galaxies described in Table 1,  
considering  the  SB effective temperature (\Ts) and ionization parameter ($U$),
 the AGN flux intensity ($F$), and  the shock velocity (\Vs) as critical parameters.

 In Sect. 2 we   present  calculations and modelling of the galaxy spectra. 
Results are presented and discussed in Sect. 3.  Concluding remarks follow in Sect. 4.

\begin{table}
\caption{Modelling  the  line ratios to \Hb=1 observed  by  van Dokkum et al. (2005) and Kriek et al. (2009)}
\tiny{
\begin{tabular}{lccccccccccc} \hline  \hline
\ line &            Obs$^1$  &     M$_{AGN}$ &      M$_{SB}$             & Obs$^2$             & m$_{AGN}$          & m$_{SB}$     \\ \hline
\ z    &            2.225    &   -           &     -       &         2.1865                    &     -              &  -   \\
\ \Ly         &        -           &       41.87       &       53.97           &    -                &  30.39             &27.7       \\
\ NV 1240     &          -           &     0.007      &     0.072              &    -                &  0.31              &1.27            \\
\ CIV 1550    &         -           &      0.53        &      0.67              &    -                &  0.8               & 2.64           \\
\ HeII 1640   &         -           &      3.43        &      3.32              &    -                &  0.89              &0.16             \\
\ [OII] 3727+ &      6.13            &    6.7          &    6.4                 & $>$5.33             & 6.                 &5.             \\
\ [NeIII] 3869+&     2.17             &    1.76        &    2.5                 &      -              &       -            &   -         \\
\ [OIII] 5007+&      10.2           &       10          &   11.                 & $<$1.36             &1.48                &1.4          \\
\ [OI] 6300+  &     $<$0.26          &      0.7        &    0.9                 &    -                &    -               &      -      \\
\ [NII] 6548+  &     2.46            &  3.2           &  3.6                    & $>$5.26                &4.7                 & 5.          \\
\  \Ha         & 3.                  &  3.            & 3.                     & 3.07                 & 3.                & 3.           \\
\ [SII] 6718  &      0.5             &    0.5       &     0.29                 & $>$1.8             & 1.1                & 1.1            \\
\ [SII] 6731  &      0.45            &    0.52      &     0.37                 & $>$0.67            &1.1                 & 1.7           \\
\ \Hb$_{obs}^3$ &  2.9 10$^{-17}$     &    -        &         -                & $<$4.2 10$^{-18}$  &     -              &      -            \\
\ \Hb$_{mod}^3$   &      -            &   0.059     &    0.051                 &    -               &0.013               & 0.009             \\
\ \Vs (\kms)    &        -           &    100       &     250                  &   -                 &180                 &150             \\
\ \n0 (\cm3)    &        -           &    150       &     130                  &   -                 &100                 & 300            \\
\ $F$ $^3$      &        -           &    40        &     -                    &-                    & 1.6               &  -           \\
\ \Ts (10$^4$ K) &       -           &    -        &     20                    &-                    &-                   & 7             \\
\ $U$            &       -          &     -        &      0.03                 &-                    &-                 & 0.007     \\
\ $D$ (10$^{19}$) cm&     -          &     2         &      1                   &-                    &1.                & 0.6    \\
\ N/H (10$^{-4}$)     &  -          &     0.8       &      0.4                  &-                   & 1.0               & 1.2  \\
\ O/H (10$^{-4}$)     &  -          &     6.6       &      5.6                  &-                   &5.6                 &6.6   \\
\  S/H (10$^{-4}$)&     -           &     0.06      &      0.06                 &-                   &0.07               &0.06    \\ \hline
\end{tabular}}

$^1$ van Dokkum et al. (2005);
$^2$ Kriek et al. (2009);
$^3$ in \erg ;
$^4$ in  10$^9$ photons cm$^{-2}$ s$^{-1}$ eV$^{-1}$ at the Lyman limit

\end{table}

\section{The models}

For the calculations of  both the  line and continuum spectra we adopted the code
  {\sc suma}\footnote{http://wise-obs.tau.ac.il/$\sim$marcel/suma/index.htm} (see also Contini  et al. 2012), which
accounts for the coupled effect of
photoionization from a radiation source and shocks. 
The input parameters related with shocks are  the  shock velocity \Vs, the atomic preshock density \n0, and
the preshock magnetic field \B0.
 \Vs determines the maximum
temperature T of the gas downstream, near the shock front. The cooling rate in the recombination zone  depends
on the square density of the gas and on T in  single slabs of gas. 
We adopt  \B0=10$^{-4}$ Gauss. 

The input parameter  that represents the radiation field  in AGNs is the power-law
flux  from the active centre $F$  in number of photons cm$^{-2}$ s$^{-1}$ eV$^{-1}$ at the Lyman limit.
The spectral indices are $\alpha_{UV}$=-1.5
and $\alpha_X$=-0.7.
For SBs the radiation is a   black body (bb). 
The input parameters are the  effective temperature of the  starburst \Ts
and  the ionization parameter $U$
(in  number of photons per  number  of electrons at the nebula).

Additional input parameters are the geometrical thickness of the emitting nebula $D$ which
 determines  whether the model is radiation-  or matter-bound, the
  abundances of  He, C, N, O, Ne, Mg, Si, S, A, and Fe relative to H and  the dust-to-gas 
($d/g$) ratio which  affect  the cooling rate.
The input parameters are all correlated throughout the Rankine-Hugoniot equations which regulate the
shock dynamics at the shock-front and downstream, and  in the ionization equations  that are used to calculate the
ion  fraction abundances for each element in each slab of gas downstream. 
Uncertainty in the calculation results is in average $<$10\%.

The primary radiation flux 
and the secondary (diffuse) radiation flux,
emitted by the gas slabs heated at relatively high temperatures by the shock,
are calculated by radiative transfer throughout the clouds. 
The  line ratios  calculated by a power-law flux  or by   bb   are different, so
we calculate  AGN and SB dominated models separately for all the galaxies.
There is no  degeneracy   for the results of  [OIII]/\Hb and  [OII]/\Hb  calculated  by SB models
as functions of   \Ts  or of $U$, because the line intensities, which
depend on the atomic configuration of the  corresponding ions, 
 show different trends with \Ts and $U$. 

Briefly,
to reproduce the data by  model calculations, we  first choose the input parameter range from the grids  
of models (Fig. 1) calculated 
by Contini \& Viegas (2001a,b, hereafter GRID1 and  GRID2, for AGN and SB, respectively).
Then we refine the fit (Tables 2-6) 
calculating  new  grids  suitable to each spectrum.
We cross-check the models resulting from  the line spectrum fit  by reproducing consistently
 the  data   in the continuum SED diagrams (Fig. 2), where
 for each model    one line  represents the
bremsstrahlung (from the gas) and  another, in the IR,  the reprocessed
radiation  from dust. An initial grain radius of 0.2 \mum is adopted.
The bb radiation  fluxes from the SB  and   from  the background old population stars
also appear in  Fig. 2 diagrams.

The results of detailed modelling  are different from those obtained by direct methods
because we calculate  the emitted line fluxes  integrating throughout the  downstream recombination region
where the  shock determines the profile of  electron temperature and density.

\begin{table*}
\caption{Modelling the  line ratios to \Hb=1 observed   by  Belli et al. (2013)}
\tiny{
\begin{tabular}{lccccccccc} \hline  \hline
\ line       & obs (A1835) & MB1$_{SB}$& MB1$_{AGN}$& obs(A773)$^1$ & MB2$_{SB}$& MB2$_{AGN}$& obs(MACS0717)&MB3$_{SB}$& MB3$_{AGN}$\\ \hline
\ z          &2.073        &     -     &     -      & 2.303         &     -     &       -    & 2.55         &      -   &     -      \\
\ [OII]3727+   & 2.17       & 2.3       & 2.2       & 4.65          & 4.6       & 4.8        & 2.27         & 2.2      & 2.3 \\
\ [NeIII]3870& 0.35        & 0.38      & 0.6        & $<$4.3        & 1         & 0.85       & $<$ 1.07     & 0.29     & 0.5 \\
\ [[OIII]5007+& $<$3.5     & 3.45      & 3.47       & 6.5           & 6.4       & 6.7        & 2.67         & 2.8      & 2.6 \\
\ \Ha        & 2.73        & 2.96      & 3.25       & 3.            & 3.6       & 3.4        & 2.9          & 2.96     & 3.16 \\
\  [NII] 6384& 0.33        &0.386       &0.39       & 0.59          & 0.47      & 0.6        & 0.7          & 0.57     & 0.59 \\
\ \Hb$_{obs}^2$ &16.6$\pm$3.1 10$^{-17}$&-          & -         &8.3$\pm$3.5 10$^{-17}$   &  -        & -       & 8.5$\pm$2.9 10$^{-17}$ &-      & -    \\
\ \Hb$_{mod}^2$ & -           &0.067      & 0.033      & -             &0.85       & 0.035      & -            & 0.017    & 0.013 \\
\ \Vs (\kms) & -           &130        & 100        & -             & 100       & 100        &  -           & 140      & 100  \\
\ \n0 (\cm3) & -           &100        & 200        & -             & 200       & 140        & -            & 100      & 80  \\
\ \Ts (10$^4$ K) & -       & 7         & -          & -             & 20        & -          & -            & 6.3      & -   \\
\ $U$           &  -       & 0.02      & -          & -             & 0.8       & -          & -            & 0.02     & -   \\
\ $F$ $^3$      & -        &  -        & 5.5        & -             & -         & 10          & -            & -        & 2 \\
\ $D$ (10$^{19}$cm)  &   -  &   1      & 1          & -             & 6         & 1          &  -           & 1        & 1.5 \\
\ N/H (10$^{-4}$)     &   -  & 0.3      & 0.3        & -             & 0.3       & 0.3        &  -           & 0.3      & 0.3 \\
\ O/H (10$^{-4}$)     &   -  & 6.6      & 5.5        & -             & 7.6       & 6.6        &  -           & 6.6      & 5.6 \\ \hline

\end{tabular}}

$^1$ reddening corrected;
$^2$  in   \erg ;
$^3$  in 10$^{9}$ photons cm$^{-2}$ s$^{-1}$ eV$^{-1}$ at the Lyman limit

\end{table*}

\section{Results and discussion}

\subsection{The physical conditions and the element abundances}

The set of  input parameters   which  yields the  best  fit  of calculated to observed data
 determines the  model for each galaxy.
It represents averaged conditions  for each object  when the   observations  cover 
the whole galaxy.

Generally,
we have found that  \Vs, \n0, and $D$  result within the ranges calculated for a large number  
of objects (Paper I, Fig. 5), while  the 
 SB temperatures (Tables 2-3-5-6)  \Ts=2.5 10$^5$K  for 
CDFS-695 (van Dokkum et al)  and A773 (Belli et al) are outstanding.
\Ts $\sim$7 10$^4$ K is found  in the quiescent galaxy 1255-0 (Kriek et al),  in  
 WISP 170-106 (Masters et al.), and in A1835 (Belli et al.) and
 \Ts=6.3 10$^4$K  in MCS 0717 (Belli et al.).

The  results  suggest that a  strong starburst activity  occurred   in
the two  sample galaxies  with maximum \Ts,  while quiescence is more adapted to
the other galaxies with \Ts $\leq$7 10$^4$ K  similar to  the starburst temperatures  calculated for most 
of the objects in Paper I and in Contini (2013).
 \Ts=2 10$^5$ K  accompanied by a relatively high \Vs in  SBs  is  suitable to young stars 
and it may indicate a revival of activity in these galaxies.  
Such high temperatures were found, for example, in  the  recurrent nova T Pyxidis
 (Contini \& Prialnik 1997), in the  symbiotic star  HM Sge at outburst (Angeloni et al. 2007), etc.
The two  `active' galaxies  also  show a relatively high $U$ ($\sim$ 1)
and  correspond to SFR$\sim$60 and 400 \Moy, respectively, which  are derived from the \Ha luminosities 
(Kennicutt 1998).

 The modelling presented in Tables 2, 3 and 6 suggests that   AGN and   SB  coexist (see also Fig. 1) 
because the line ratios in each spectrum are satisfactorily fitted (within the observed errors) by 
both AGN and SB models.
The same  was found  throughout many local merger galaxies, due to  
mutual  feedbacks.
The   AGN fluxes  calculated for the present  sample are similar to those found in 
low luminosity AGN (LLAGN), except for  J1000+0221S (Amor\'{i}n et al. 2014)
which   suits a Seyfert 2 galaxy.

We have found  O/H  close to solar  (6.6 10$^{-4}$, Allen 1976) for all the objects (Fig. 3). 
van Dokkum et al refer to diagnostic diagrams calculated by solar O/H and Kriek et al.   
mention  solar or supersolar  O/H.
N/H modelled for Kriek et al. quiescent galaxy is similar to  the  maximum  (N/H=1.4 10$^{-4}$)
calculated in local galaxies (e.g. NGC 7212, Contini et al. 2012, fig. 14).
 Masters et al  suggest an average O/H=2.18 10$^{-4}$ for their sample  objects, while our models
 show O/H  slightly higher than solar (Allen 1976), but within  the  solar  
O/H= 8.5 10$^{-4}$ given by Anders \& Grevesse (1989)(cf Paper I, table 1). 
Calculated N/H are generally lower than solar ($\sim$ 10$^{-4}$), in agreement with the results
obtained for galaxies in the same z range.
For J1000+0.221S  we  found solar O/H and N/H, while
Amor\'{i}n et al. derived O/H=2.75$^{2.88}_{2.6}$ 10$^{-5}$  
by the strong-line and calibration method presented by Maiolino et al. (2008) (see Paper I).

\begin{figure}
\centering
\includegraphics[width=4.4cm]{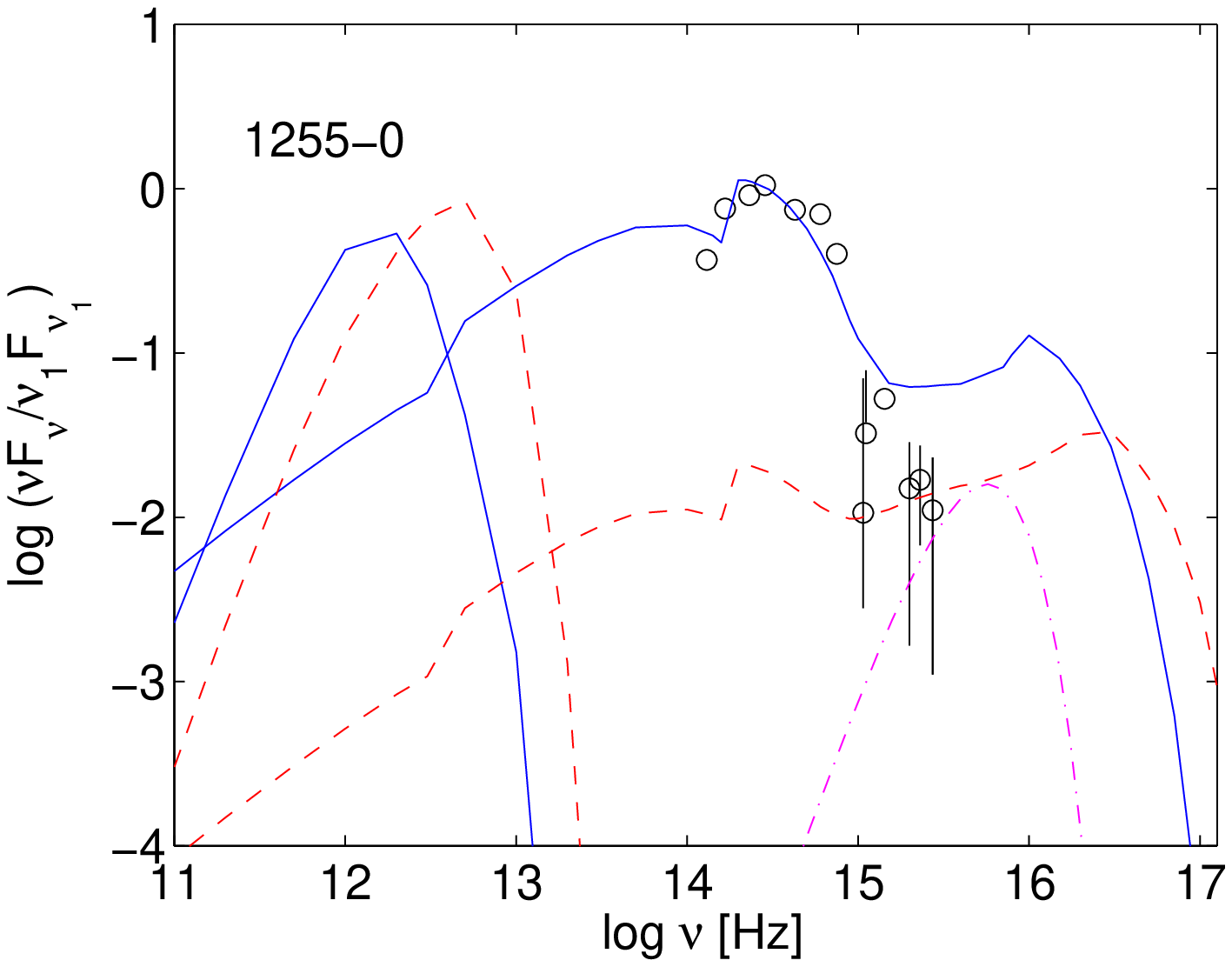}
\includegraphics[width=4.4cm]{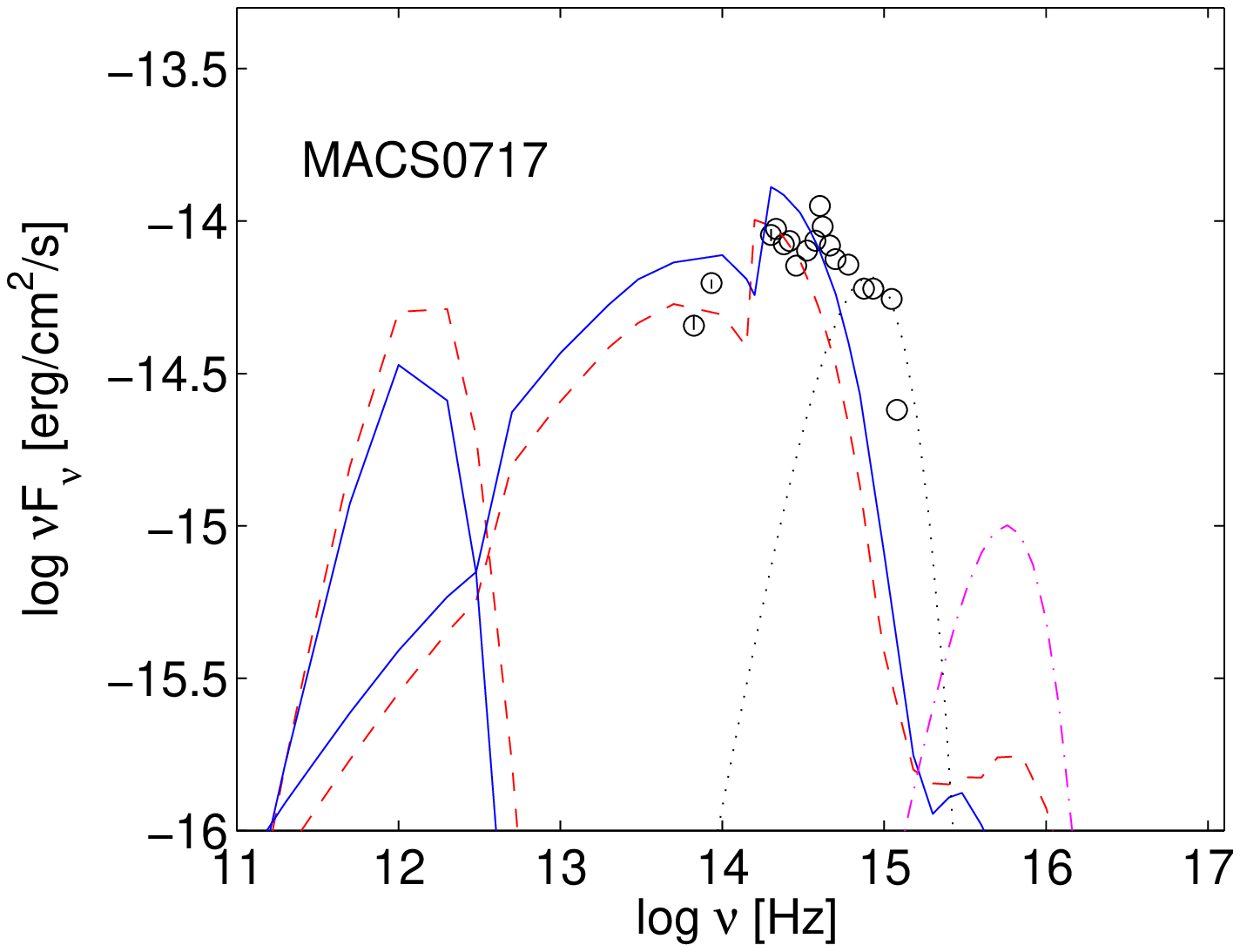}
\caption{SED of Kriek et al (2009, fig. 2) and Belli et al (2009, fig. 1) MACS0717 galaxies.
Black circles : the data;
blue solid lines : AGN models;
red dashed lines : SB models;
magenta dot-dashed lines : bb radiation calculated for \Ts ;  black dotted : bb for \Ts=10$^4$K}
\end{figure}

\begin{figure}
\centering
\includegraphics[width=6.0cm]{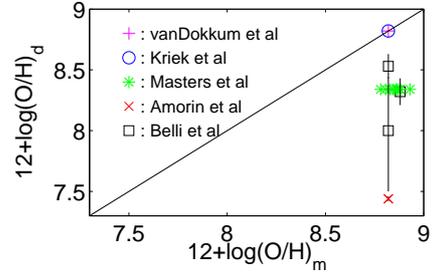}
\caption{Direct method (d) versus model (m) results for O/H} 
\end{figure}

\subsection{Notes on individual galaxies}

{\bf CDFS-695 }. We  cross-checked the results obtained by   modelling the line spectrum (Table 2, col.3 and 4) 
calculating the distance r of the emitting cloud from the AGN in the NLR.
The \Hb flux observed at Earth (2.9 10$^{-17}$ \erg)  after reddening correction (Osterbrock 1974, eq 7.6, fig. 7.1)
results \Hb $_{obs}$= 6.10$^{-16}$ \erg. Combining it with the \Hb flux calculated at the cloud (\Hb$_{mod}$) :
\noindent
 \Hb$_{obs}$ d$^2$ = \Hb$_{mod}$ r$^2$   (where d is the distance to Earth) 
 and adopting a filling factor $\sim$ 1 throughout the NLR,
 r results $\sim$ 1.8 kpc, in rough  agreement with van Dokkum et al who suggested the presence of  shocks 
on the basis of 
the line ratios extending to  $\sim$ 10 kpc from the nucleus.
The distance of the emitting clouds from the SB region,  calculated in the same way,
results r$<$ 1.9  kpc,  adopting a filling factor $<$ 1.
The SB region does not  cover the entire galaxy. 
SB have typical radii of 100-1000 pc (1-10\% of the size of their 'host' galaxies).

{\bf 1255-0}. 
We constrained the  models (Table 2) 
of this  `quiescent' galaxy  (Kriek et al 2009) 
by the  relatively low [OIII]5007+/[OII]3727+,
 [OII]/\Hb, and [OIII]+/\Hb (Table 2, col. 5). 
The  modelling  results appear in Table 2, col. 6 and 7 (m$_{AGN}$, m$_{SB}$).
The relatively low calculated  \Ha flux ($\sim$  3 \Hb$_{mod}$) partly  explains the low observed \Ha 
flux (1.29$^{+0.16}_{-0.18}$  $\times$ 10$^{-17}$ \erg) and 
luminosity \La(0.46$^{+0.06}_{-0.06}$ $\times$ 10$^{-42}$ erg s$^{-1}$)  confirming 
 the low SFR   observed by Kriek et al.

In Fig. 2 (left diagram)   we  model the  continuum SED   presented
by Kriek et al. (2009, fig. 2). The data are normalized to 
unity at 1 \mum. 
 The  bremsstrahlung calculated
by model m$_{AGN}$ reproduces the data  for $\nu$ between 10$^{14}$ and 10$^{15}$ Hz.
The  Planck function corresponding to 7 10$^4$ K
 peaks in the UV where the flux is  strongly absorbed by the Galaxy.
A few datapoints in the SED between 10$^{15}$ and 10$^{16}$ Hz fit
the   bremsstrahlung calculated by m$_{SB}$ with perhaps some contribution from
 the bb flux from the SB.
Models m$_{AGN}$  and m$_{SB}$ were calculated by    dust-to-gas ratios (by number) 
of 10$^{-15}$ and  10$^{-14.3}$, respectively.

{\bf A1835, A773, and MACS0717}. The results for the galaxies 
 from the Belli et al. sample are presented in Table 3.
Although the observed line ratios are typical of SBs, we have  searched for
AGNs in each object. 
The results   show that  AGN could be present (Fig. 1).
 \Ts  is relatively high, in particular for A773 (2 10$^5$ K). 
The emitting clouds are  geometrically thick ($D$$\geq$ 10$^{19}$ cm)  which  seems unusual for AGNs  
in  this z range ($D$$\leq$ 10$^{16}$ cm, Contini 2013, fig. 3).
The models selected for MACS0717 (Table 3)  are cross-checked by fitting  the  SED in Fig. 2. 
The observed photometry
follows the bremsstrahlung maximum. The peak frequency depends on \Vs and the peak intensity on $F$. 
The old star background population  contributes to
the MACS0717 SED by a bb flux corresponding to a temperature of 10$^4$ K.

\begin{table}
\centering
\caption{Modelling  the  line ratios (\Hb=1) observed by Masters et al (2014)}
\tiny{
\begin{tabular}{cccccccccccccccc} \hline  \hline
\ WISP   &   z   &  [OII]       &  [OIII]      & \Ha   & [NII]     & \Hb$_{obs}$  $^1$ \\ 
\        &       &  3727+       & 5007+        &       & 6583      &                 \\ \hline
\ 173-205& 1.444 & 1.19         & 6.27         & 2.97  & -         & 7.5(0.4) 10$^{-18}$\\
\  m1    &   -   & 1.1          & 6.3          & 3.03  & 0.19      &-  \\
\ 9-73   & 1.454 & 1.88         & 5.3          &4.     & 0.51      & 31.6(1.3) 10$^{-18}$\\\
\  m2    & -     & 2.0          & 5.4          & 3.2   & 0.40      &-  \\
\ 25-53  & 1.486 & 2.13         & 2.55         & 2.8   & 0.51      & 31.0(0.9) 10$^{-18}$\\\
\  m3    & -     & 2.0          & 2.4          & 3.    & 0.42      & -\\
\ 46-75  & 1.504 & 2.8          & 1.36         & 3.05  & 0.45      & 6.6(1.1) 10$^{-18}$\\\
\  m4    & -     & 2.7          & 1.46         & 3.    & 0.37      &- \\
\ 170-106& 2.165 & 3.57         & 8.64         & 4.    & $<$0.35   & 3.7(0.4) 10$^{-18}$\\\
\  m5    & -     & 3.           & 8.32         & 2.9   & 0.13      &- \\
\ 64-210 & 2.177 & 1.49         & 6.76         & 4.2   & 0.75      & 20.1(1.4) 10$^{-18}$\\\
\  m6    & -     & 1.4          & 6.3          & 2.9   & 0.43     &- \\
\ 27-95  & 2.192 & 2.69         & 4.87         & 4.17  & 0.52     & 18.6(0.7) 10$^{-18}$\\\
\  m7    & -     & 2.67         & 4.57         & 3.1   & 0.4      &-\\
\ 90-58  & 2.212 & 2.15         & 2.65         & 3.79  & 0.54     & 8.7(0.3) 10$^{-18}$\\\
\  m8    & -     & 2.1          & 2.65         & 3.    & 0.5      & -\\
\ 56-210 & 2.304 & 1.55         & 6.25         & 2.8   & -        & 2.7(0.4) 10$^{-18}$\\\
\ m9     &  -    & 1.6          & 6.67         & 3.1   & -        &-\\
\ 206-261& 2.315 & 1.08         & 6.64         & 4.1   & $<$0.15  & 7.4(0.5) 10$^{-18}$\\\
\ m10    & -     & 1.1          & 6.67         & 3.    & 0.13 \\ \hline
\end{tabular}}

$^1$ \Hb observed flux in  \erg. The uncertainty appears in parenthesis

\end{table}

\begin{table}
\centering
\caption{Gas physical conditions and relative abundances calculated for the Masters et al. (2014) sample}
\tiny{
\begin{tabular}{lcccccccl} \hline  \hline
\  mod &  \Vs & \n0  & $U$  & \Ts       & $D$          & N/H        & O/H      & \Hb$_{mod}$   \\
\         & \kms & \cm3 & -    & 10$^4$ K  & unit$^1$      & 10$^{-4}$  & 10$^{-4}$ &unit$^2$       \\ \hline
\   m1    & 180  & 160  & 0.07 & 4.8       & 0.6          & 0.8        & 7.3      & 0.0036            \\
\   m2    & 170  & 120  & 0.06 & 4.2       & 0.5          & 1.5        & 6.8      & 0.001             \\
\   m3    & 140  & 150  & 0.03 & 4.0       & 1.0          & 0.7        & 7.       & 0.0046          \\
\  m4     & 140  & 150  & 0.01 & 4.5       & 1.0          & 0.3        & 6.       & 0.0043          \\
\   m5    & 320  & 200  & 0.12 & 7.5       & 3.0          & 0.1        & 8.5      & 0.1            \\
\   m6    & 250  & 210  & 0.1  & 5.6       & 2.0          & 0.7        & 7.5      & 0.065           \\
\   m7    & 230  & 100  & 0.05& 4.        & 1.0          & 0.85       & 7.6      & 0.0013         \\
\   m8    & 200  & 100  & 0.04 & 4.        & 2.0          & 0.8        & 6.4      & 0.0042        \\
\   m9    & 180  & 160  & 0.07 & 4.5       & 0.4          & 0.8        & 7.3      & 0.0018         \\
\   m10   & 180  & 160  & 0.07 & 4.8       & 0.6          & 0.6        & 7.6      & 0.0035          \\ \hline
\end{tabular}}

 $^1$ 10$^{16}$ cm ;
 $^2$  \erg ;

\end{table}

{\bf The WISP galaxies}.
The modelling  of  line ratios  from  the Masters et al sample of galaxies (Table 1)  
with  prominent \Ha and/or [OIII]
lines, is shown  in Table 4.
 We  selected the galaxies corresponding to 2.8 $<$\Ha/\Hb$<$ 4.2,
i.e. those which do not need  reddening correction 
(Osterbrock 1974).  Modelling results  appear in Table 5.
 WISP 170-106  shows a relatively high  shock velocity  (320 \kms)  and a relatively high $U$ (0.12) for \Ts=7.5 10$^4$ K
 suggesting  a  fading activity.

{\bf J1000+0221S}.  Amor\'{i}n et al. (2014) report the spectrum observed from  this strongly 
gravitationally lensed galaxy.
The  observed line ratios  appear in Table 6.
They are  similar to those   calculated by models presented in  GRID1 (table 4, mod 36) for AGNs and in GRID2 
(table 1, mod 6  and table 2, mod 7) for SBs.
The AGN flux $F$ =3 10$^{11}$
photons cm$^{-2}$ s$^{-1}$ eV$^{-1}$ at the Lyman limit, is suitable to  Seyfert 2 AGN and
is higher by  $\sim$7-200  than  in the other galaxies  of the present sample.
 Amor\'{i}n et al. define this object as a  metal-poor star forming galaxy  based on the 
low O/H and low $U$ deduced by empirical +photoionization method.
Our  models  show different results because 
of the coupled effect of photoionization and shocks. This would  lead to
a different  mass-metallicity- SFR relation.

\begin{table}
\centering
\caption{Comparison of model results with the data for J1000+0221S
(Amor\'{i}n et al. 2014)}
\tiny{
\begin{tabular}{lcccccccc} \hline  \hline
\                        & obs        & M$_{AGN}$ & M1$_{SB}$ & M2$_{SB}$ \\ \hline
\        z               & 3.417      &     -     &     -     &       -     \\
\ [OII]3727              &$<$0.3      & 0.3        &$<$0.5    &0.43     \\
\ [NeIII]3869           & $<$0.2     & 0.2        & 0.25     & 0.4    \\
\ [OIII]5007+4959        & 5.9$\pm$2.6& 4.7        &5.5       &8.45     \\
\ \Hb                    & 1          & 1          & 1        & 1      \\
\ \Hb$_{mod}$ $^1$       & -          & 21.9       &0.13      & 2.58   \\
\ \Vs (\kms)             & -          & 200        &100        & 200  \\
\ \n0  (\cm3)            & -          & 200        &100        & 200   \\
\ $F$  $^2$                & -          & 3          & -         & -     \\
\ \Ts(10$^4$K)           & -          & -          & 5         & 5    \\
\     $U$                & -          & -          &0.1         & 1    \\
\ $D$ (10$^{19}$ cm)    & -          & 1          & 1         & 1    \\ 
\   O/H (10$^{-4}$)      &  -           & 6.6        & 6.6      & 6.6  \\
\ N/H  (10$^{-4}$)       & -          &0.9          &  0.9     & 0.9  \\ \hline
\end{tabular}}

$^1$ in \erg ; 
$^2$ in 10$^{11}$ photons cm$^{-2}$ s$^{-1}$ eV$^{-1}$ at the Lyman limit

\end{table}

\section{Concluding remarks}

The  high  starburst temperature  (\Ts$\sim$ 2 10$^5$ K)  
suggesting activity  in  the two galaxies CDFS-695 (van Dokkum et al. 2005) and  A773 (Belli et al 2013)
is not seen in the  other galaxies of the modelled samples. 
Activity in terms of a  Seyfert 2  AGN  flux is found in
the J1000+0221S galaxy (Amor\'{i}n et al 2014).
The ionization parameter  $U$ calculated for the galaxy bulk 
 shows  larger scattering than $F$
because $U$ depends on the hot source radiation flux  as well as on  the ISM  conditions
So it is less significant to activity than
$F$ which depends  mainly on the intensity of the  non-thermal radiation
flux from the AGN. 

By detailed modelling of the spectra we have found  O/H  close to solar  (Allen 1976, 
Anders \& Grevesse 1989). 
N/H are generally lower than solar, in agreement with the results
obtained for galaxies at those redshifts.

\section*{Aknowledgements}
I am grateful to the referee for useful comments.

\section*{References}

\def\ref{\par\noindent\hangindent 18pt}
\tiny{
\ref Allen, C.W. 1976 Astrophysical Quantities, London: Athlone (3rd edition)
\ref Amor\'{i}n, R. et al. 2014, ApJL, 788, L4
\ref Anders E., Grevesse N. 1989, Geochim. Cosmochim. Acta, 53, 197
\ref Angeloni, R., Contini, M., Ciroi, S., Rafanelli, P. 2007, AJ, 134, 205
\ref Belli, S., Jones, T., Ellis, R.S., Richard, J. 2013 ApJ, 772, 141
\ref Contini, M. 2014, A\&A, 564, 19, Paper I
\ref Contini, M. 2013, arXiv1310.5447
\ref Contini, M., Cracco, V., Ciroi, S., La Mura, G. 2012, A\&A, 545, 72
\ref Contini, M. \& Prialnik, D. 1997, ApJ, 475, 803
\ref Contini, M. \& Viegas, S.M.A. 2001a, ApJS, 132, 211, GRID1
\ref Contini, M. \& Viegas, S.M.A. 2001b, ApJS, 137, 75, GRID2
\ref Kennicutt, R.C. 1998, ARA\&A, 36, 189
\ref Kriek, M., van Dokkum, P.G., Labbe', I., Franx, M., Illingworth, G.D., 
Marchesini, D., Quadri, R.F. 2009, ApJ, 700, 221
\ref Kriek, M. et al. 2006, ApJ, 649, L71
\ref Maiolino, R. et al. 2008, A\&A, 488, 463
\ref Masters, D. et al. 2014,  ApJ, 785, 153
\ref Osterbrock, D.E. 1974 in  Astrophysics of Gaseous Nebulae. W.H.Freeman and Company,
San Francisco, Ed. 
\ref van Dokkum, P.G., Kriek, M., Rodgers, B., Franx, M., Puxley, P. 2005, ApJ, 622, L16
}
\end{document}